\def\wh{wormhole }
\def\whs{wormholes }
\def\beq{\begin{equation}}
\def\eeq{\end{equation}}
\def\bea{\begin{eqnarray}}
\def\eea{\end{eqnarray}}

\documentstyle[aps,floats]{revtex}

\begin{document}

\input epsf
\renewcommand{\topfraction}{1.0}
\twocolumn[\hsize\textwidth\columnwidth\hsize\csname
@twocolumnfalse\endcsname

\title{Might some Gamma Ray Bursts be an observable signature of
natural wormholes?}

\author{Diego F.Torres${^{1,2,}}$
Gustavo E. Romero${^{2,3,4}}$ and Luis A. Anchordoqui${^{2}}$}

\address{${^1}$ Astronomy Centre, CPES,
University of Sussex, Falmer, Brighton BN1 9QJ United Kingdom}
\address{${^2}$Departamento de F\'{\i}sica,  Universidad Nacional
de La Plata, C.C. 67, 1900 La Plata, Argentina}
\address{${^3}$ Instituto Argentino de Radioastronom\'{\i}a,
C.C. 5, 1894 Villa
Elisa, Argentina}
\address{${^4}$ Instituto Astron\^omico e Geof\'{\i}sico, USP. Av. M. Stefano 4200,
CEP 04301-904, S\~ao Paulo, SP, Brazil}

\maketitle

\begin{abstract}
The extragalactic microlensing scenario for natural wormholes is examined.
It is shown that the main
features of wormhole lensing events upon the light of distant Active Galactic
Nuclei (AGNs) are similar to some types of
already observed Gamma Ray
Bursts (GRBs). Using
recent satellite data on GRBs, an upper limit to the negative mass density
-- ${\cal O}\,(10^{-36})$ g cm$^{-3}$ -- under the form of wormhole-like
objects is presented.\\
{\it PACS number(s): 98.62.Sb, 04.20.Gz} \hspace{5.2cm} SUSSEX-AST-98/1-2
\end{abstract}

\vskip2pc]

\section{Introduction}

Ten years after the seminal paper by Morris and
Thorne \cite{motho}, we face the following situation:
there is no observational
evidence supporting the existence of natural wormholes nor
serious theoretical reasons for its impossibility \cite{VISSER-BOOK}.
Black holes shared such a status during years until the discovery of galactic
X-ray sources and quasars in the 1960s.
Wormholes, entities that warp spacetime in
such way as to provide shortcuts to separated regions of the universe
or even a way
to allow a backward time travel, require the violation of the energy
conditions (technically speaking, the null energy condition) in order to
exist.
The energy conditions are conjectures
that are widely used to prove issues concerning singularities
and black hole thermodynamics; they constitute just plausible
statements, like the positivity of the energy density. However, several
situations in which the energy conditions are violated
are known; perhaps the most quoted of them is the
Casimir effect. These violations
are typically very small (of order $\hbar$)
and it is far from clear
whether there could be macroscopic quantities
of this kind of {\it exotic} matter. Nevertheless, there is nothing
really compelling to prevent its occurrence
and wormholes might naturally exist
\cite{6-visser,7-visser}.

Very recently,
the consequences of the energy conditions
were confronted with possible values of the Hubble parameter
and the gravitational redshifts of the oldest stars in the galactic halo
\cite{VISSER-HUBBLE}. It was
deduced that for the currently favored values of $H_0$,
the strong energy condition should have been violated sometime between the
formation of the oldest stars and the present epoch.
On the other hand, negative gravitational masses
(underdensities in the primordial universe)
have been proposed as an
explanation of the voids observed in the extragalactic space \cite{PIRAN-0}.
An early universe cosmic network of wormholes has also been
suggested as an alternative solution for the cosmological
horizon problem \cite{Hochberg}.
Mann \cite{MANN} have found, in addition, that dense regions of negative mass
can undergo gravitational collapse,
ending up in exotic black holes that could populate the universe
contributing to the bulk of total dark matter.
All these works clearly show that it is at least possible that
natural wormholes or other negative mass objects might exist. Then
the study of their possible observational effects
deserves serious consideration.
Although no universal mechanism to generate a relic density of
exotic matter is well established at present (because of our ignorance of
quantum gravity laws), several interesting ideas have been
recently proposed in the literature, like, for instance, the enlargement
of \wh throats --via inflation--
from the quantum foam to macroscopic sizes \cite{ROMAN}.
Despite current theoretical speculations
suggest that the existence of compact objects of negative mass
is plausible, their amount has not been yet constrained by
observations.
To provide such a constraint is the main goal of this paper.

As far as we are aware, the first observational proposal to search for natural
wormholes or similar gravitational negative anomalous compact objects was
presented by Cramer et al. \cite{CRAMER}  (see also Ref. \cite{DIAZ}).
They suggested that gravitational
microlensing effects of these objects upon the light of background stars
could produce
MACHO\footnote{MACHO: massive compact halo object.}-like events
\cite{PACHINSKY-ANNU-REVIEW}, although with different
(asymmetric) temporal profiles.
Partial
analysis of the results of several ongoing microlensing monitoring programs
seems to show that wormhole-like objects are not present in the
dark halo of our galaxy. (Hereafter,
when speaking of negative masses, we shall think in this ingredient
as always threading a wormhole. Although this can be relaxed
for the development and analysis of the ideas to be considered, we shall
do it just
because it can provide useful numerical estimates and a pretty theoretical
framework).

In this paper we shall study the microlensing
scenario for an extragalactic natural wormhole
acting upon light coming from an Active Galactic Nucleus (AGN).
It will be shown that such anomalous microlensing event
would produce lightcurves very similar to some already observed Gamma Ray
Bursts (GRBs) \cite{BATSE-ANNU-REVIEW} and that
this can be used to constrain the amount of negative
mass in the universe. Preliminary results on this issue were introduced in
\cite{CONCURSO} and briefly commented on in \cite{NS}.

The paper is ordered as follows. The next section will review
the relevant observational characteristics of the GRB phenomenon. Sec. III
will deal with the negative mass lensing formalism. Afterwards, we shall
analyze the consequences of negative-mass microlensing with an AGN as
background source in Sec. IV.
The possible nature of the lenses
is treated in Sec. V, while the BATSE database is briefly discussed in the
Sec. VI.
The final two sections deal with the cosmological consequences of
a negative mass distribution of compact objects and the conclusions.

\section{Gamma Ray Bursts}

Gamma ray bursts are flashes of high energy radiation that can be brighter,
during their brief existence, than any other gamma ray source in the sky. The
bursts present an amazing variety of temporal profiles, spectra, and
timescales
that have puzzled astrophysicists for almost three decades
\cite{BATSE-ANNU-REVIEW}.
In recent years, our observational insight of this phenomenon has been
dramatically
increased by the huge amount of data collected by the {\it
Burst and Transient Source
Experiment} (BATSE) on board the {\it Compton Gamma Ray Observatory}
(CGRO), a satellite launched by NASA in 1991.
BATSE observations have confirmed that no
large clustering or anisotropies are present in the sky distribution of GRBs
(see Ref. \cite{BATSE-ANNU-REVIEW} and references therein).
We shall give a brief account of the most relevant characteristics
of GRBs
below.

\begin{itemize}

\item {\bf Temporal Profile:} The temporal distribution of the bursts is one
of the most striking signatures of the GRB phenomenon.
There are at least four classes of distributions,
from single-peaked bursts, including the fast rise
and exponential decaying FREDs, their inverse or anti-FREDs
to chaotic structures.
There are well separated
episodes of emission, as BATSE
triggers \# 1235 or \# 222 and bursts with
extremely complex profiles, as \# 160 or \# 404. Most of the bursts are time
asymmetric but some are symmetric, as \# 408.

\item {\bf Timescales:} Burst timescales go through the 30{\rm ms}
scale to hundreds of seconds. The measurement of these timescales
is a rather complicated task, since it may depend on the intensity
of both the background and the source. At high energies ($>100$  MeV),
some extremely long bursts have been detected. For instance, GRB
940217 showed a high energy photon
($\simeq $ 20 {\rm GeV}) 1.5 {\rm hours} after
the bulk of the detection.

\item {\bf Spectra:} A unique and common characteristic of
GRBs is that most of
their power is received in energies higher than 50 {\rm KeV}. Their
spectrum approximately
follows a power law $N(E) \propto E^{-\beta}$, with $\beta \in (1.7,2.7)$.
It is interesting to note that there is
no correlation between the
spectral index and the morphology of the temporal
profile or the location in the sky.

\end{itemize}

A special issue relevant for the ideas to be presented is
the possible repetition of the bursts.
Before BATSE was launched, repetition was analyzed by Schaefer and
Cline \cite{Schaefer}, who provided two timescales for repetition
corresponding
to monoluminous or multiluminous sources. More recently, Quashnock and Lamb
\cite{QL} found that a significant fraction of the GRBs in BATSE 1B
catalogue could repeat over timescales of months. They found that many GRBs
are grouped within angular scales
smaller than 4$^o$, which is the mean error in position of BATSE detections
\cite{BATSE-CATALOG}.
However, a similar statistical technique was used by
Narayan and Piran \cite{PIRAN} to prove
that there are also an important fraction
of GRBs with antipodal positions,
thus suggesting that any statistical bias
or selection could produce both effects.
Other preliminary tests made by Petrosian and Efron
\cite{Petrosian} and Strohmeyer et al. \cite{Stro}
suggested
that there are some repetition in the sample
with timescales of years in
about 20$\%$ of the bursts at most.
It was also pointed out that a failure in the CGRO tape recorders
could have hidden some repeating sources \cite{BATSE-ANNU-REVIEW}.

More recently, another statistical work concluded that the number of repeated
bursts cannot be larger than $7\%$ of the sample \cite{Hakkila}. The most
recent and complete repetition study on the BATSE catalogue has been
carried out by
Tegmark et al. \cite{TEGMARK}. They analyzed the angular power spectrum of
1122 GRBs finding that no more than 5 \% can be labeled as
repeaters at the 99 \% confidence level. By now, evidence for repetition
is very suggestive but,
perhaps, not compelling. This point might
be clarified with forthcoming technologies, especially when
detectors with improved spatial resolution become available and studies on
individual GRB repetition can be made unambiguously.

The isotropic distribution of GRBs strongly suggests an
extragalactic origin which has been
recently confirmed by the direct measurement
of high-redshifted absorption lines of the
optical counterpart of the GRB 970508
\cite{metzger}. If the sources are so far, the energy necessary to
produce the observed events by an intrinsic mechanism is astonishing: about
$10^{51}$ erg of gamma rays
must be released in less than 1
second \cite{PIRAN-3}.\footnote{The observed flux $F$ of a source at a
distance $d$ is related with the intrinsic luminosity $L$ by $L=4\pi d^2 F$.}
The most popular model to date to
produce such an event is the merger of two compact stars (two neutron stars or
a neutron star and a black hole) in a distant galaxy.
As a result of the merging,
a relativistic expanding fireball is formed. It is believed that the
interaction of the blast with the surrounding medium produces lower energy
(X-ray, optical, may be radio) counterparts of the original GRB; again, the
reader is referred to \cite{BATSE-ANNU-REVIEW} for a survey of the current
literature.
This fireball paradigm, however, is not free of problems, as can be seen, for
instance, in Ref.  \cite{dar}.

The wide variety of burst
profiles, the statistical evidence for
GRB repetition and some spectral properties remain
unexplained by an unique, consistent model.
There is such a large variety of individual events that every model proposed
has to face a large number of counter-examples.
These facts are suggesting, perhaps, that the origin of such a
complex phenomenon might have more than one explanation.
In fact, this idea was recently proposed in \cite{grqc}, where it was concluded
that two or more distinct groups
of bursts, probably of different origin, could be classified.

Ten years ago McBreen and Metcalfe \cite{MCBREEN} proposed that GRBs could be
due to microlensing of background AGNs. At that time there was no direct
observational evidence for that AGNs, like quasars and BL Lac objects,
were strong gamma ray emitters, which is now a well
established fact \cite{mattox}. However, their
model was ruled out due to the fact that most of GRBs are time-asymmetric
\cite{nemi}, which is incompatible with microlensing by ordinary matter.
In spite of this, gamma
ray emitting AGNs can be outstanding background sources for producing extrinsic
GRB-like events if their radiation is gravitationally focused on the observer.
This focusing effect must be provided by
interposed lenses made of exotic, negative mass matter
which can thread, for instance, a wormhole structure.
As we shall see, individual amplification events
are not necessarily time-symmetric in such a case,
and repetition can occur as a
consequence of different caustic crossing within some source-lens-observer
configuration. Moreover, the expected high energy spectral features and the
lower energy manifestations of the phenomenon are strikingly similar to some
already observed GRB events.

\section{Negative mass lensing}

We shall briefly review now some concepts of gravitational lensing by
negative masses. The assumed geometry is that of an extragalactic wormhole of
negative mass $-M$ crossing with velocity $V$ the line of sight to some
distant AGN. We shall follow the presentation given by Cramer et al.
in Ref. \cite{CRAMER} but we shall take into account the extragalactic
nature of the lensing.

The Einstein radius of a negative mass is given by

\beq
\label{rae}
R_e=\left( \frac{4GMD}{c^2} \right)^{ 1/2} ,
\eeq
where, aside from the usual meaning of the constants $c$ and $G$,
$D$ represents an
effective lens distance. This is a model-dependent parameter;
in particular, it varies for different values of
$H_0$ and $\Omega_0$, the Hubble constant and the
energy density parameter at the present
time, respectively.  The general expression for $D$ is

\beq
D=\frac{D_{ol} D_{ls}}{D_{os}},
\eeq
where $D_{ol}$, $D_{ls}$ and $D_{os}$ are the observer-lens, lens-source and
observer-source angular diameter distances, all them computed as in
\cite{NARAYAN}

\beq
D(z_i,z_j)=\frac {2c}{H_0} \frac{\left( 1-\Omega _0 - G_i G_j \right)
\left( G_i -G_j \right) }{\Omega _0^2\left( 1+z_i \right) \left( 1+z_j
\right)^2 },
\eeq
with

\beq
G_{i,j}=\left( 1 + \Omega _0 z_{i,j} \right)^{\frac 12},
\eeq
and $z_i$ the cosmological redshift of the object $i$.

The variability timescale $T$ of a microlensing event is defined as the time
that takes the line of sight to the source
to cross the Einstein radius of the lens: $T=R_e/V$.
The overall relative intensity ${\rm I}_{{\rm neg}}$ is the modulation in
brightness
of the background source detected by the observer. This is given by
\cite{CRAMER}

\beq
{\rm I}_{{\rm neg}}=\frac { B^2 -2}{B \sqrt{B^2 -4} },
\eeq
where

\beq
B(t)=B_0 \left( 1 + \left( \frac {t}{t_v} \right)^2 \right)^{1/2}.
\eeq
Here, $B_0$ is the time-dependent dimensionless impact parameter and $t_v$
is the transit time across the distance of the minimum impact parameter, $t_v
\propto T$. Taking ${\rm I}_{{\rm neg}}=0$ for $|B|<2$, it is possible
to obtain the light enhancement profile for a negative amount of mass $M$.
These curves, see Fig. 1, must be divided in
two groups. For $B_0>2$, the light profiles are similar to the positive mass
cases but provide larger light enhancement than that given by a similar
amount of positive mass. For $B_0<2$, curves are sharper
and present brief light enhancement; they have
divergences (caustics) of the intensity and then an
immediate drop to zero.
This happens at two given times, solutions of $B^2-4=0$;
thus, for time running from $-\infty$ to $+\infty$, and during the same
microlensing event, we obtain two divergences
and two drops, of specular character. This is
seen by the observer as two bursting events separated by a time $\sim T$.
Unlike the $B_0>2$ case, these individual bursts
present light profiles asymmetric under time reversal.

\begin{figure}
\label{Ineg-fig} 
\centering 
\leavevmode \epsfysize=6cm \epsfbox{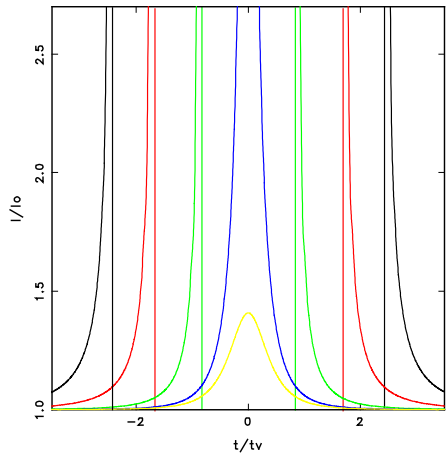}\\ 
\caption{Overall relative intensity 
${\rm I}_{{\rm neg}}$ for a microlensing event 
provided by negative amounts of matter. From 
the corners of the graph towards the centre, 
the curves corresponds to $B_0=$0.5, 0.75,
1, 2 and 2.2.}
\end{figure} 

A critical requirement for such a microlensing event to
occur is that the size of
the background source projected onto the lens plane must not be larger
than the
Einstein ring of the lensing mass \cite{chang}. Otherwise, light from outside
the Einstein ring would smoother out the gravitationally induced variability.
Background sources whose size is a fraction of the
Einstein radius are then amplified by significant factors, while sources whose
projected sizes largely exceed the Einstein radius are negligibly amplified.
Since AGNs have emission regions of different scales for different radiation
wavelengths, the spectrum of an observed microlensing event will depend on the
mass of the lens as well as on the involved redshifts.

Finally, it is important to note that a point source is amplified by
an infinite amount at a caustic crossing, but any physical extent leads to
finite amplifications \cite{SCHNEIDER-BOOK}. This {\it point-mass}
infinity  also happens in the Einstein ring of a positive point mass
configuration.

\section{AGNs as background sources}

AGNs are compact extragalactic sources of extraordinary luminosity. They can
radiate as much energy per unit of time as hundreds of normal galaxies.
Most of
this energy comes, however, from a region much smaller than the mean distance
between two stars in our galaxy. Although
AGNs emit across the entire electromagnetic spectrum,
from radio to gamma rays, recent
observations by two instruments on board the CGRO, the {\it Compton Telescope}
(COMPTEL) and the {\it Energetic Gamma Ray Experiment Telescope} (EGRET),
have shown that many of these objects radiate most of their power
in the form of gamma rays
(see, for instance, Ref. \cite{monti} and references therein).

In spite of the existence of many classes of AGNs like quasars, BL Lac objects,
Seyferts, and others, it is widely accepted that the same basic mechanism
operates in  all of them. This standard working paradigm of AGNs assumes
that the
central engine powering these objects is  a supermassive ($\sim 10^8$
M$_{\odot}$) black hole + accretion disk system. Energy is generated by
gravitational infall of material which is heated to very high temperatures in
the dissipative, optically and geometrically thick disk. Along the rotational
axis of the system two jets of ultrarelativistic electron-positron plasma are
collimated by a yet not well-established mechanism. The emission signatures of
these jets, however, are clearly detected by interferometric radio observations
in many objects \cite{Begelman}. The several classes of AGNs are usually
interpreted as a viewing effect due to the basic anisotropy of the
phenomenon.

The inner part of the accretion disk emits X-rays. The origin of
this X-rays is
not quite clear, but it is commonly thought that the UV/optical continuum
emission from the disk is up-scattered in energy by inverse-Compton
scattering off relativistic electrons in a hot corona surrounding the
disk, a process
referred to in the literature as `Comptonization' of the input (UV/optical)
spectrum \cite{Kafatos}. A purely thermal origin, however, cannot be completely
ruled out. Anyway, the jets must traverse throughout this external
radiation field. Inverse Compton interactions between the relativistic leptons
that form the jets and the ambient photons produce gamma rays which,
due to the
relativistic bulk motion of the source, are Doppler enhanced in the beam
direction. Probably, the accretion disk is not the only source of seed X-ray
photons, being these additionally produced in the jet itself by
synchrotron
emission and also reprocessed in a surrounding halo of dense clouds (see Ref.
\cite{blan} for details). Anyway, the compactness of the observed gamma ray
source will be limited by
gamma ray absorption in the UV-X-ray ambient field due to
pair production.

The requirement that the pair production opacity to infinity equals unity
naturally defines gamma-spheres of radius $r_{\gamma}$ for a given gamma ray
energy $E_{\gamma}$ \cite{blan,becker}: no gamma rays with energy higher than
$E_{\gamma}$ can be observed from radii smaller than $r_{\gamma}$ because they
would be absorbed by pair production. Notice that the sizes of the successive
gamma-spheres increase for increasing energies of the observed photons. This
fact has important consequences for gravitational lensing: the high energy
spectrum will be differentially amplified, presenting a cutoff at energies for
which
the size scales of the gamma-spheres exceed the Einstein ring of the lens. The
observer should see a gamma ray burst when the line of sight to the AGN
intersects a caustic, with a similar spectrum to the original source at lower
energies (keV to MeV) but with a cutoff at higher energies (GeV to TeV).
AGN's high energy spectra are well represented by power a law
$F(E_{\gamma})\propto E_{\gamma}^{-\alpha}$, with $\alpha$ in the range
1.5-3.0
\cite{monti}, remarkably similar to many GRBs detected by BATSE and EGRET.
It is
also interesting to notice that high energy continuum spectra of GRBs present a
cutoff at energies of a few GeV \cite{BATSE-ANNU-REVIEW}.

Optical emission is originated by synchrotron mechanism in the jets of AGNs. The
optical region can be coextensive with the outer gamma-spheres \cite{blan} and,
due to the acromaticity of gravitational light bending, simultaneous or
quasi-simultaneous optical bursts can be expected for a microlensing event. Due
to the larger size of the emitting region, the optical flare will have also
larger timescales than
those associated to the inner gamma-spheres. Radio emission, instead, is
originated far down the jet, in regions where the plasma density is considerably
lower (around 1 parsec from the black hole). In most cases such sizes
might exceed the Einstein radius of potential small microlenses leading to
GRB-like events without counterparts at radio wavelengths.

Summing up, the central region of AGNs is a suitable background source for
microlensing by compact extragalactic lenses. The resulting events, if some
of the lenses are wormhole-type objects, should very much resemble
GRBs: brief flares of gamma rays, with power law continuum spectra and
X-ray to optical counterparts, in many cases on larger timescales.
The event-averaged high energy spectrum is remarkably similar to a
typical AGN-spectrum. Moreover, the total duration of some
extremely large events, including the detection of very high energy photons
at the end, could be exactly what one would expect from microlensing:
since the most energetic gamma-spheres are the bigger ones,
their crossing time must be larger. In other cases, like GRB 970111,
no X-ray or
optical emission have been detected despite the bursts were well in the
field of
view of very sensitive instruments like the
Beppo-SAX satellite. This fact can be a
straightforward consequence of the relatively large sizes of the
corresponding
emitting regions when compared with the inner gamma-spheres.

\section{Wormholes as negative mass lenses}

In order to get a feeling of the involved magnitudes
in a \wh microlensing event
let us consider a concrete example. We shall focus on the model assuming
$H_0=100\,{\rm km \, s^{-1}\, Mpc^{-1}}$, $\Omega_0=1$ and a
cross velocity for the extragalactic lens equal to 5000 km s$^{-1}$.
Defining $D=( 2c/H_0 ) {\cal D}$, we find

\beq
R_e=1.04 \times 10^{12} \left( \frac{M}{M_\odot} \right)^{1/2}
{\cal D}^{1/2}\; {\rm km}.
\eeq
Table \ref{TABLE1} shows some negative masses required
to get a particular variability timescale $T$,
for a given configuration of lens-source redshifts,
along with the corresponding Einstein radius. Due to the size constraints
on the AGN's emitting region previously mentioned (Sec. III), we have

\beq
x\leq 2 R_e \frac{D_{os}}{D_{ol}},
\eeq
where $x$ is the linear size of the emitting region in the core of the AGN.
Replacing $R_e$ we get a constraint over the mass of the possible lenses:

\beq
M\geq \frac{c^2 x^2}{16\, G} \; \frac{D_{ol}}{D_{os}D_{ls}}.
\eeq

Considering that the typical size of the gamma-spheres for energies of
$\sim$ 1
GeV are $x\sim 10^{-3}$ parsecs \cite{becker}, we find in the case of the
redshifts quoted in Table 1 that $|M| \geq 1.27\times10^{-3} M_\odot$, and
thus
one should expect that burst repetition due to crossing of the two
different
caustics in a single event should take  several months, even for substellar
wormhole masses.

\begin{table}
\caption{Typical negative masses for a lens-source redshift
configuration given by $z_l=0.25$ and $z_s=2.5$ in order to provide
each of the variability timescales. The corresponding Einstein radii are also 
shown. It was assumed a low extragalactic velocity equal to 
1000{\rm km s$^{-1}$}.}

\begin{tabular}{ccc}
 $T$ {\rm years}  & $-M/M_\odot$ & $R_e$ {\rm km} \\ \hline
1/12  & $ 9.68 \times 10^{-5}$ & $2.59 \times 10^9$  \\
1/2  & $ 3.50 \times 10^{-3}$ & $1.55 \times 10^{10}$  \\
1  &  0.014  & $3.15 \times 10^{10}$  \\
10  &  1.430  & $3.15 \times 10^{11}$  \\
\end{tabular}
\label{TABLE1}
\end{table}

We shall show now that these stellar and
substellar masses are attainable with a simple
\wh configuration. We shall use a well known example of \wh geometry, the
absurdly benign wormhole, introduced
in \cite{motho}. This special kind of \wh is a solution of the Einstein
field equations corresponding to the metric

\beq
ds^2 =  - dt^2 + \left( 1- \frac {b(r)}{r} \right) ^{-1} dr^2 +
r^2 d\Omega_2^2,
\label{metric}
\eeq
with

\bea
b(r) &  = &  b_0 \left( 1- \frac{r-b_0}{a_0} \right)^2, \hspace{1cm}
 {\rm if} \hspace{0.2cm} b_0 \leq r \leq b_0+a_0 ,  \label{b-def1} \\
b(r) &  = & 0,   \hspace{1cm}
{\rm if} \hspace{0.2cm} r > b_0+a_0  . \label{b-def2}
\eea
In this solution, $b_0$ is the throat radius and $a_0$ is a cut-off in the
energy density; space-time is empty for $r > b_0+a_0$.
The timelike field equation is

\beq
\label{f-set-1}
\rho=\frac{b^\prime}{8 \pi G r^2}.
\eeq

\mbox{} From (\ref{f-set-1}) we can integrate for $b(r)$ and define a mass
function by

\beq
\label{mass}
b(r)=b(r_0) + \int _{r_0}^r 8 \pi G \rho r^2 dr \equiv  2Gm(r),
\eeq
which yields the total mass of the wormhole \cite{VISSER-BOOK},

\beq
\frac{M}{M_\odot} = 0.337\; b_0 \left( 1-
\left( \frac{b_0}{a_0} \right)^2 \right) .
\eeq
The numerical factor arises from the use of solar mass units
while the radius is given in {\rm km}.
Note that this mass is not necessarily negative,
and it depends on the relationship between the values of $b_0$ and $a_0$.
This does not mean that null energy condition (NEC) is not violated,
because
some of the other two inequalities of this condition need to fail:
there are no wormholes fulfilling NEC.
Table \ref{TABLE2} presents some illustrative numerical
values of the parameters in this simple configuration.
However, it is important to remark that this solution is by no means
special; other geometries
can account for stellar-size masses,  without
being spherically symmetric \cite{VISSER-PRD-R}.

\begin{table}
\caption{Masses for the absurdly bening wormhole.}
\begin{tabular}{ccc}
$b_0$ ({\rm km}) & $a_0$ ({\rm km})& $-M/M_\odot$ \\\hline
1  & 0.50  & 1.011\\
5  & 4.99 &  $1.34 \times 10^{-3}$ \\ 
\end{tabular}
\label{TABLE2}
\end{table}

Since \whs connect two otherwise separated asymptotic regions, there are two
asymptotic masses which can, in general, differ. Exchange of matter
between both wormhole mouths can modify their mass ratio starting a
process that could lead to a large (stellar-size) negative mass in one of
the mouths \cite{VISSER-BOOK,CRAMER}.
We expect that the computation of masses with (\ref{mass}) will be
possible whenever the stress-energy tensor is confined to some fixed
radius, in such a way that spacetime
becomes vacuum and described by a piece of Schwarzschild solution.

\section{On BATSE detections}

Distinctive features of \wh microlensing are repetition of the event and a
definite asymmetry in the profiles of the repeaters: the initial bursts are
anti-FREDs whereas their counterparts are FREDs. A study of the temporal
asymmetry in the BATSE database, then, can be useful to enlighten the role
played, if any, by microlensing in the production of GRBs.

The time asymmetry of a GRB lightcurve can be quantitatively estimated
using the third moment of the time profile given by

\beq
{\cal A}=\frac{<(t-<t>)^3>}{<(t-<t>)^2>^{3/2}},\label{A}
\eeq
where the brackets denote average over all data weighted with the number
of counts. For a time symmetric burst results ${\cal A}=0$, while those
bursts
with faster rises (falls) than falls (rises) present ${\cal A}>0$ (${\cal A}
<0$). A
determination of ${\cal A}$ for a sample of 631 bursts from BATSE 3B
catalogue
\cite{LINK} shows that  32 \% of the profiles present ${\cal A}<0$. This
result clearly means that {\it microlensing by wormholes cannot be the
only physical mechanism behind the GRB phenomenon}. Since fireballs
naturally account for short rising times they are the best candidate to
explain most of the events. However, the formation of the fireball
requires a sudden release of energy which is radiatively dissipated during
the blast wave expansion and, consequently, GRBs with ${\cal A}<<0$ remain
unexplained.

Bursts with ${\cal A}<0$ cannot be directly considered as tracers of
wormholes because not all of them repeat. As we have mentioned in Sect.
II, the whole sample is consistent just with a 5 \% of repetition. This
means that just about 56 out of 1122 events in the BATSE 3B catalogue
could be originated by wormholes. The identification of these individual
bursts cannot be made unambiguously because of the large positional error
boxes of BATSE measurements. We shall discuss here some candidates.

Fig. 2 shows BATSE
trigger \# 257 (GRB 910602 \cite{BATSE-CATALOG})
which was detected a couple of months after the
start of the space mission. It is a typical, single-profile, clearly asymmetric
GRB. Its duration was $\sim 80$ s, with a peak flux of $\sim1.7$ photons
cm$^{-2}$ s$^{-1}$. This kind of event could be produced by a
single wormhole microlensing occurrence with dimensionless impact parameter
$B_0<2$. In Fig. 2, we have
superimposed to the observational lightcurve, a theoretical
microlensing curve. The timescale of the microlensing event is
$T\sim3.2$ years.

\begin{figure}
\label{fig2} 
\centering 
\leavevmode\epsfysize=6cm \epsfbox{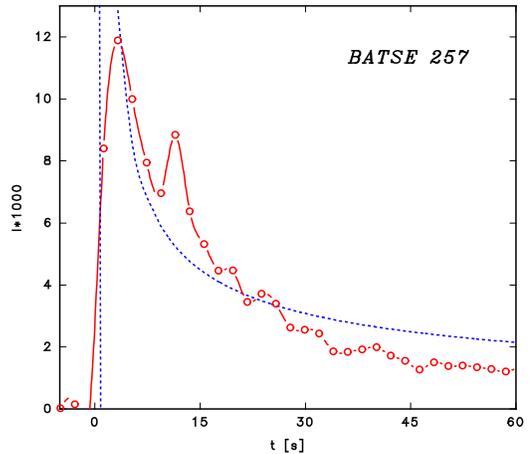}\\ 
\caption{This is BATSE Trigger \#257.
The hollow circle points represent the summed 
number of counts of the total number of 
channels of the two triggered detectors (det. 1 and 2.). 
The dotted curve is a theoretical curve
${\rm I}_{{\rm neg}}$ with the following values: $t_v=1.09\times 10^{8} {\rm 
s}$, 
${\rm I}_0=2.81$. The position of the theoretical caustic is 1.63 {\rm s}.}
\end{figure} 

Since the event asymmetry corresponds to a second-caustic
crossing, a similar event with opposite asymmetry, corresponding to
first-caustic passing, should have occurred in April 1988, three years before
CGRO launching. Assuming the set of redshifts and velocity
mentioned in Table 1, we find that
a wormhole of mass $M\sim\; -0.146 \, M_\odot$ might have been responsible for the
event.
Similar results may be obtained for other triggers.

Some remarkable GRBs in the database, from the point of view of
wormhole microlensing, are BATSE triggers
$\#1653$ (June 17 1992,
at $l^{II} = 131.18^\circ$, $b^{II} =-41.25^\circ$)
and $\#2110$ (December 30 1992, at $l^{II} = 132.91^\circ$,
$b^{II} =-42.87^\circ$). A symmetry analysis using (\ref{A}) shows that,
when the background is subtracted in such a way that the peak structure is
emphasized over that of the surrounding foothills, the bursts present an
anti-FRED--FRED structure \cite{bennett}. In addition, the position error
boxes for
these GRBs include three AGNs, namely PG 0117+213, 0109+200, and the BL
Lac object 0109+224. This particular object has been detected at high
energies by ROSAT and other satellites
\cite{della} and there are many reports of its fast variability at
different wavelengths \cite{valtaoja}. It is, consequently, an excelent
candidate for background source in a \wh microlensing event.
Unfortunately, its redshift is unknown at present so we cannot infer from
the event timescale ($T\sim6.5$ months) a range of possible masses for the
lens. If a redshift $z=1.5$ is assumed and the wormhole is halfway, then
the mass of the \wh results $-0.12 M_{\odot}$.
Calculations, however, are not very sensitive to $z$.

\section{Cosmological consequences}

At this stage it would
be worth obtaining an estimate of an upper
limit to the amount of negative mass that could exist in the universe.
With this aim, we shall assume that the negative matter is under
the form of wormhole-like compact objects, and we
shall estimate the optical depth considering
that GRB repeaters detected by BATSE were caused by them.
This will provide a consistent upper limit on the possible
number of {\it isolated} wormholes in the universe.
Wormholes linked to galactic
halos are not taken into account in this calculation and
should, instead, be treated in the way
described by Cramer et al. in Ref \cite{CRAMER}.

The concept of optical depth was originally introduced in the context of
gravitational lensing by Vietri and Ostriker \cite{vietri}, and it was applied
by Paczy\'nski \cite{pac} to the problem of gravitational microlensing by
objects belonging to the dark halo of our own galaxy. The optical depth to
microlensing can be defined as the fraction of solid angle covered with Einstein
rings of the lensing objects. If it is smaller than unity (which is certainly
the case when wormholes are considered as lenses) it provides a measure of the
probablity of microlensing. The total optical depth due to all lenses placed
between the background source and the observer is given by

\beq
\tau=\frac{4\pi G}{c^2}D_{os}^2\;\int^1_0  |\rho(x)|\, x \,(1-x) \, dx
\eeq
where $\rho$ stands for the mass density distribution of negative matter under
the form of wormholes and $x\equiv D_{ol}/ D_{os}$,
\cite{PACHINSKY-ANNU-REVIEW}. Clearly, the value of $\tau$
depends on the model adopted for the distribution of lensing matter along the
line of sight towards the distant sources. For simplicity, we shall
adopt here a constant density. Then,

\beq
\tau= \frac{2 \pi}{3}  \frac{G\, D_{os}^2 \, |\rho|}{c^2}.
\eeq
$|\rho|$ is expected to be extremely small, otherwise cosmological effects
concerning a wormhole-filled universe should be evident. Then, $\tau$, the
probability of detecting a microlensing event onto a given background source, is
almost negligible. Fortunately, the number of background AGNs seems to be huge:
about ten percent of the objects detected in the Hubble Deep Field images are of
this class \cite{almaini}. This makes the total number of
potential background
sources for microlensing by wormholes as high as $10^9$.
The number of events observed in a lapse $\Delta T$ is

\beq
N=\frac{2n}{\pi} \tau \frac{\Delta t}{T},
\eeq
where $n$ is the total number of
background AGNs and $T$ is a typical timescale for microlensing,
\cite{pac}.
Then, using both previous
formulae in favor of $|\rho|$, we get

\beq
\label{negrho}
|\rho|=\frac 34 \frac{T}{\Delta t}
\frac Nn \frac {c^2}{G} \frac {1}{D_{os}^2}.
\eeq
In (\ref{negrho}) there are quantities of two different kinds.
Most of the magnitudes involved are related to observation. We have in this
group the already mentioned number of background sources and the observed
number of BATSE triggers that may be associated with repetition,
$N=1122 \times 5/100$ during $\Delta t =3$ {\rm years}
of operation. The angular diameter distance
of the source is also fixed because cosmological distribution of AGNs
seems to peak somewhere between $z_s=2$ and $z_s=3$, and so we can adopt
an intermediate value of $z_s=2.5$. On the other hand, we have one
model-dependent magnitude, the variability timescale of the problem, $T$.
As $T \simeq R_e /V$, we note that both, the mass and the velocity of the
lens, are degrees of freedom of (\ref{negrho}). As we want to find
an upper bound on $|\rho|$ we shall choose a conservative extragalactic
velocity of 5000 {\rm km s$^{-1}$}. Regarding the mass, we saw in
the previous section that
a mass of $-0.1 M_\odot$ seems to fit an observed BATSE trigger and is
suitable for timescales from months to years, consistently
with GRB-repetition intervals. In the absence of any other clue respect to
possible masses of natural wormholes we adopt this value. In the calculation
we also take into account
the fact that one wormhole should produce two GRBs of the sample.
With these figures, we obtain

\beq
\label{number}
| \rho |\leq 9.05 \times 10^{-36} \,{\rm g\, cm ^{-3}}.
\eeq

\mbox{} The mass density (\ref{number}) must be considered as a large
upper bound on the possible amount of negative matter in the universe.
Clearly, this amount is too small to produce significant
cosmological consequences. For comparison,
we recall that
a lower limit for the mass contribution due to galaxies in the universe is $6
\times
10^{-31}$ {\rm g cm$^{-3}$}, and the critical
density is of order
$1.9 \times 10^{-29}$ {\rm g cm$^{-3}$} (see, for instance, Ref.
\cite{COLES}).

\section{Conclusions}

We have shown that microlensing events produced by wormholes with AGNs as
background sources very much resemble certain types of GRBs; types that
cannot be explained in standard models.
We then used observational data on GRBs to determine
an upper limit for the amount of wormhole-like objects in the universe.
This upper limit is enough to see that negative matter hardly would have any
influence in cosmology.
An unusual feature of the presented scenario is that, while GRB
repetition has previously been seen as a strong evidence for noncosmological
origin, the microlensing model accepts it warmly: sources are
cosmological and repetitions arise from different caustic crossings.
This model implies that not only
some bursting events must repeat, but also that they should
do it with temporal profiles of specular character. This makes the model
capable to be falsified.
We expect that, with the improvement of the observational
techniques and the increase of the GRB sample,
more exact limits to the amount of the negative mass will be available.
Forthcoming technologies and satellites such as
the {\it High Energy Transient Explorer} (HETE),
the next {\it Gamma Ray Large Area Space Telescope} (GLAST)
and the current Beppo-SAX satellite
will help to improve burst position measurements yielding light onto
the repetition phenomenon.

Whether the laws of physics, in some deep realization,
forbid the violations of the energy conditions in the large amount
needed to produce stellar-size compact objects of negative matter,
is something not yet clear. But, if the universe does
admit wormholes geometries in it, it is very likely that some of the
GRBs may be caused by a
microlensing mechanism, being this one of the main conclusions of this work.
As an immediate spinoff, we have
the converse fact, i.e. that if there were no possible
burst in a large, perhaps not already obtained sample,
which could be associated with wormhole-like lensing, then it should
be understood as a serious objection to the existence
of anomalous compact objects in the universe.

\acknowledgments

It is a pleasure to thank S. E. Perez-Bergliaffa,
Z. Abraham, A. R. Liddle and J. Lidsey for comments and
critical readings of the manuscript.
We are also indebted to B. Link and R. Epstein for providing us
their results on symmetry analysis. This research has made use of the
NASA/IPAC Extragalactic Database, which is operated by the Jet Propulsion
Laboratory, California Institute of Technology, under contract with the
National Aeronautics and Space Administration.
Our work has been partially supported by the Argentine agencies CONICET
(D.F.T. and G.E.R) and  ANPCT (PMT-PICT 0388) (G.E.R),
the Brazilian agency FAPESP (G.E.R), the British Council (D.F.T),
Fundaci\'on Antorchas (D.F.T) and the FOMEC program (L.A.A.).

\end{document}